\documentclass[aps,prd,onecolumn,eqsecnum,amsmath,nofootinbib,preprintnumbers]{revtex4}%

\usepackage{color,graphicx,float,subfigure}
\usepackage{amsfonts,amssymb,theorem,mathrsfs,times}
\usepackage{bm}
\usepackage{mathtools}
\usepackage{amsfonts,amssymb,theorem,mathrsfs}
\usepackage{dsfont}
\usepackage{setspace}
\usepackage{graphicx}

{\theorembodyfont{\upshape}
}
{\theorembodyfont{\upshape}
}
{\theorembodyfont{\upshape}
}
{\theorembodyfont{\upshape}
}
{\theorembodyfont{\upshape}
}
{\theorembodyfont{\upshape}
}

\newcommand{\dalm}{\kern1pt\vbox{\hrule height 0.9pt\hbox{\vrule width
0.9pt\hskip 2.5pt\vbox{\vskip 5.5pt}\hskip 3pt\vrule width
0.3pt}\hrule height 0.3pt}\kern1pt}

\begin{document}
\preprint{\hfill {\small {ICTS-USTC/PCFT-21-37}}}
\title{A Bound on the Rate of  Bondi Mass Loss}

%

\author{ Li-Ming Cao$^{a\, ,b}$\footnote{e-mail
address: caolm@ustc.edu.cn}}

\author{ Long-Yue Li$^b$\footnote{e-mail
address: lily26@mail.ustc.edu.cn}}

\author{ Liang-Bi Wu$^b$\footnote{e-mail
address: liangbi@mail.ustc.edu.cn}}

\affiliation{$^a$Peng Huanwu Center for Fundamental Theory, Hefei, Anhui 230026, China}

\affiliation{${}^b$
Interdisciplinary Center for Theoretical Study and Department of Modern Physics,\\
University of Science and Technology of China, Hefei, Anhui 230026,
China}


\date{\today}

\begin{abstract}
To ensure the light (emitted far away from the source of gravity) can arrive at the null infinity of an asymptotically flat spacetime, it is shown that the rate of  Bondi mass aspect has to satisfy some conditions.
In Einstein gravity theory, we find the sufficient condition implies a bound on the Bondi mass $m$, i.e., $|\dot{m}|\leqslant 0.3820~c^3/G$. This provides a new perspective on  Dyson's maximum luminosity.
However, in Brans-Dicke theory, the sufficient condition depends on the behavior of the radiation field of the scalar.
Specifically, the photons can escape to the null infinity when the scalar gravitational radiation is not too large and the mass loss is not too fast.
\end{abstract}


\maketitle


\section{Introduction}
Gravitational radiation has attracted wide attention since LIGO found gravitational waves from the merger of a binary black hole in 2016~\cite{LIGOScientific:2016aoc}.
At present, we detect the gravitational waves  far away from the source. So it is important to study the asymptotic structure of the spacetimes which are endowed with gravitational waves.
Actually, in 1960s, Bondi {\it et. al.} and Sachs have established a suitable formalism to study the behavior of asymptotically flat spacetimes in general relativity (GR)~\cite{Bondi:1960jsa,Bondi:1962px,Sachs:1962wk}.
By using the metric assumed by Bondi {\it et. al.} and Sachs, it is known that the gravitational radiation of an asymptotically flat spacetime results in mass loss~\cite{Bondi:1960jsa,Bondi:1962px, Sachs:1962wk}, see also a  pedagogical review~\cite{Madler:2016xju}.
This of course implies that gravitational waves really carry energy.

In vacuum, gravitational waves affect the behavior of electromagnetic waves. The amplitudes of electromagnetic waves and gravitational waves both propagating in the same direction
are oscillating in the Minkowski background, which means the conversion between photons and gravitons~\cite{Frolov:1998wf}.
If the initial directions of gravitational waves and electromagnetic waves are perpendicular, electromagnetic waves change direction and flow at an angle to the initial waves~\cite{Patel:2021cat}.
So it is vital to explore the impacts of gravitational waves on electromagnetic waves. In some realistic models, comparing to the gravitational waves, the electromagnetic waves have very large frequency.
This implies that the geometric optics approximation is suitable for the electromagnetic waves, and one can consider a simpler model in which light propagates (or the null geodesics) on  the spacetime with gravitational waves.

Recently,  Amo, Izumi, Tomikawa, Yoshino, and Shiromizu have investigated the behavior of the null geodesics near future null infinity in an asymptotically flat spacetime~\cite{Amo:2021gcn}.
They solved the geodesic equations  and found the dependence of the coordinate $u, r$ with respect to an affine parameter $\lambda$.
By this, they got a sufficient condition that the photons emitted outwards at large $r$  can arrive at the null infinity, and it is found that  in four dimension there is some possibility that the photons can not reach the null infinity.
However,  the physical meaning of the sufficient condition is not clear up to date.
Based on the analysis in~\cite{Amo:2021gcn}, we further study the sufficient condition and find that condition naturally imposes a constraint on the rate of the Bondi mass loss, i.e., $\dot{m}$.
Actually, it gives a range on $\dot{m}$, whose absolute value has a upper bound $0.3820c^3/G$.
In a word,  the photons can escape to the infinity implies that the mass loss is slower than the absolute value of this bound.
After multiplying the abstract value of the bound by $c^2$, we get a luminosity for the asymptotically flat system. This provides a clue of the existence of the maximum luminosity proposed by Dyson long time ago~\cite{Dayson1963,Barrow:2017atq}.

In GR, the lower bound of $\dot{m}$ is independent of any features of the asymptotically flat spacetime. So it is reasonable to ask a question whether the lower bound of the rate of the mass loss shares the same property in other gravity theories,
especially for the gravity theories with nontrivial scalar degrees of freedom.
To answer the question,  we investigate the behavior of the null geodesics in Brans-Dicke theory (BD).  Brans-Dicke theory is one typical example of a scalar-tensor theory, a class of theories in which there is a scalar field
coupling to gravity nonminimally. It can be  found  from $f(R)$ theory by an appropriate conformal transformation~\cite{Clifton:2011jh, DeFelice:2010aj}. Recent years, the mass loss and memory effect in BD are investigated in~\cite{Hou:2020tnd,Tahura:2020vsa}.
The coordinate $r$ chosen by Tahura and her collaborators corresponds to the determinant of the metric of angular part~\cite{Tahura:2020vsa}. This selection on $r$ is the same as the one in GR.
In this case, the components of Riemann tensor approach to zero when $r$ becomes large, so the spacetime is asymptotically flat.
The coordinate $r$ chosen by Hou and Zhu is different from the one in~\cite{Tahura:2020vsa} by a factor given by the scalar field~\cite{Hou:2020tnd}.
Although the spacetimes are asymptotically flat both in Einstein frame and Jordan frame, the $g_{uu}$ approaching to $-1$ at infinity
will bring a lot of convenience in calculation, so we will discuss the null geodesics in the coordinates used by Hou and Zhu.
Based on this procedure, we find the bound of $\dot{m}$ in BD depends on the radiation field of the scalar.

This paper is organized as follows. We will give a brief review on the Bondi-Sachs formalism of Brans-Dicke theory in asymptotically flat spacetime in Section \ref{B1}.
In Section \ref{B2}, we study the asymptotic behavior of null geodesics and study whether the photons can arrive at the null infinity.
The condition that $r$ increases with the affine parameter $\lambda$ is given in Section \ref{B3}.
In Section \ref{B4}, we get the sufficient condition that the photons can arrive at the null infinity.
The result is discussed in Section \ref{B5}, and we  find it implies that the rate of the Bondi mass loss has a lower bound. 
Section \ref{B6} is devoted to the case in GR. This subsection will account for the reason why the upper bound of $|\dot{m}|$ is $0.3820~c^3/G$ in GR.
In Section \ref{B7}, spherically symmetric spacetimes are discussed, and one can clearly find the effect of the scalar field on the photons. In Section \ref{ConDis}, we give a conclusion and discussion.
The relation between our bound and the Dyson's maximum luminosity is discussed there.


\section{Brans-Dicke Theory}\label{B1}

In this section, we will give a brief review on the Bondi-Sachs formalism of the asymptotically flat spacetime in BD. The action of BD in the Jordan frame  without matter field is given by~\cite{Brans:1961sx}
\begin{equation}
S=\frac{1}{16 \pi G} \int \mathrm{d}^{4} x \sqrt{-g}\left(\varphi R-\frac{\omega}{\varphi} \nabla_{a} \varphi \nabla^{a} \varphi\right)\, ,
\end{equation}
where $R$ is the Ricci scalar, $g$ is the determinant of the metric, $G$ is the gravitational constant, $\varphi$ is a scalar field, and $\omega$ is a positive constant. The equations of motion are given by
\begin{equation}
\label{A1}
R_{a b}-\frac{1}{2} g_{a b} R = \frac{8 \pi G}{\varphi}\mathcal{T}_{a b}\,,
\end{equation}
and
\begin{equation}
\label{A2}
\nabla_{c} \nabla^{c} \varphi =0\, .
\end{equation}
Here, $R_{ab}$ is the Ricci tensor, and $\mathcal{T}_{a b}$ is the effective stress-energy tensor which is given by
\begin{equation}
\label{A16}
\mathcal{T}_{a b}=\frac{1}{8 \pi G}\left[\frac{\omega}{\varphi}\left(\nabla_{a} \varphi \nabla_{b} \varphi-\frac{1}{2} g_{a b} \nabla_{c} \varphi \nabla^{c} \varphi\right)+\nabla_{a} \nabla_{b} \varphi-g_{a b} \nabla_{c} \nabla^{c} \varphi\right]\, .
\end{equation}
In the following discussion, we set $G=c=1$, and will restore them when necessary. In the Bondi-Sachs formalism~\cite{Barnich:2010eb}, the metric has a form
\begin{equation}
\mathrm{d} s^{2}=e^{2 \beta} \frac{V}{r} \mathrm{d} u^{2}-2 e^{2 \beta} \mathrm{d} u \mathrm{d} r+h_{A B}\left(\mathrm{d} x^{A}-U^{A} \mathrm{d} u\right)\left(\mathrm{d} x^{B}-U^{B} \mathrm{d} u\right)\, ,
\end{equation}
where $A=2,3$, $x^2=\theta, x^3=\phi$, and $\beta$, $V$, $U^A$, and $h_{AB}$ are six metric functions which depend on all of the coordinates.
In asymptotically flat spacetime, the determinant condition is~\cite{Hou:2020tnd}
\begin{equation}
\label{A3}
\det \left(h_{A B}\right)=r^{4}\left(\frac{\varphi_{0}}{\varphi}\right)^{2} \sin ^{2} \theta\, ,
\end{equation}
and the coordinate $r$ is defined by this condition in some sense.
The expansions of $\varphi$ and $h_{AB}$ can be written as
\begin{equation}
\label{expansionvarphi}
\varphi=\varphi_{0}+\frac{\varphi_{1}}{r}+\frac{\varphi_{2}}{r^{2}}+\mathcal{O}\left(\frac{1}{r^{3}}\right)\, ,
\end{equation}
and
\begin{equation}
\label{expansionh}
h_{A B}=r^{2} q_{A B}+r c_{A B}+d_{A B}+\mathcal{O}\left(\frac{1}{r}\right)\, ,
\end{equation}
where $q_{AB}$ is the standard metric of the 2 dimensional unit sphere and its determinant is given by $q=\sin^2\theta$. Based on the condition (\ref{A3}), one can define two traceless symmetric tensors $\hat{c}_{AB}$ and $\hat{d}_{AB}$ as
\begin{equation}
c_{A B}=\hat{c}_{A B}-q_{A B} \frac{\varphi_{1}}{\varphi_{0}}\, ,
\end{equation}
and
\begin{equation}
d_{A B}=\hat{d}_{A B}+q_{A B}\left(\frac{1}{4} \hat{c}_{CD} \hat{c}^{CD}+\frac{\varphi_{1}^{2}}{\varphi_{0}^{2}}-\frac{\varphi_{2}}{\varphi_{0}}\right)\, .
\end{equation}
From Eqs.(\ref{A3}), (\ref{expansionvarphi}), and (\ref{expansionh}), it is not hard to shown that
\begin{equation}
\hat{c}^A{}_A=q^{AB}\hat{c}_{AB}=0\, ,\qquad
\hat{d}^A{}_A=q^{AB}\hat{d}_{AB}=0\, .
\end{equation}
The tensor $\hat{c}_{AB}$ also satisfies the identity for tensors on the 2d sphere~\cite{Godazgar:2018vmm}, i.e.,
\begin{equation}
\hat{c}_{AC}\hat{c}_B{}^C=\frac{1}{2}\hat{c}_{CD}\hat{c}^{CD}q_{AB}\, .
\end{equation}
By using the equations of motion Eq. (\ref{A1}) and (\ref{A2}), we can get the fall-off behaviors of the functions $\beta$, $V$, and $U^A$. They can be expanded as
\begin{eqnarray}
\label{VbetaU}
V&=&-r+2M+\mathcal{O}\left(\frac{1}{r}\right)\, ,\nonumber\\
\beta &= &-\frac{\varphi_1}{2\varphi_0}\frac{1}{r}+\left[-\frac{\hat{c}_{AB} \hat{c}^{AB}}{32}+\frac{1-2 \omega}{16}\left(\frac{\varphi_{1}}{\varphi_{0}}\right)^{2}-\frac{\varphi_{2}}{2 \varphi_{0}}\right]\frac{1}{r^2}+\mathcal{O}\left(\frac{1}{r^{3}}\right)\, ,\nonumber\\
U^A &=&-\frac{\eth_B\hat{c}^{AB}}{2}\frac{1}{r^2}+\left(-\frac{2}{3} N^{A}+\frac{1}{3} \hat{c}^{A B} \eth_{C} \hat{c}^{C}{}_{B}\right)\frac{1}{r^{3}}+\mathcal{O}\left(\frac{1}{r^{4}}\right)\, ,
\end{eqnarray}
where the function $M$ is the aspect of the Bondi mass, and $``\cdot"$ denotes  the derivative with respect to the coordinate $u$, and
$$N_{AB}=-\partial_u \hat{c}_{AB}$$
is the so-called Bondi news tensor~\cite{Ashtekar:1981hw}, and $N^A{}_B=q^{AC}N_{BC}$. $N^A$ is the angular momentum aspect, and the symbol $\eth$ is the covariant derivative which is  compatible with the metric $q_{AB}$. By using the equations of motion,  the evolution of the function $M$ is given by
\begin{equation}
\label{A4}
\dot{M}=-\frac{1}{4} \eth_A \eth_B N^{A B}-\frac{1}{8} N_{A B} N^{A B}-\frac{2 \omega+3}{4}\left(\frac{\dot{\varphi}_1}{\varphi_{0}}\right)^{2}\, ,
\end{equation}
and $\dot{\varphi}_1$ corresponds to the scalar aspect of gravitational wave~\cite{Hou:2020tnd}.
From equations (\ref{VbetaU}), it is easy to find the nontrivial components of the metric
\begin{eqnarray}
g_{uu}&=&-1+\frac{2M+\varphi_1/\varphi_0}{r}+\mathcal{O}\left(\frac{1}{r^2}\right)\, ,\nonumber\\
g_{ur}&=&-1+\frac{\varphi_1}{\varphi_0r}+\mathcal{O}\left(\frac{1}{r^2}\right)\, ,\nonumber\\
g_{uA}&=&\frac{\eth_B \hat{c}^B{}_A}{2}+\mathcal{O}\left(\frac{1}{r}\right)\, ,\nonumber\\
g_{AB}&=&r^2 q_{AB}+r c_{AB} +\mathcal{O}\left(1\right)\, .
\end{eqnarray}
These will be used in the calculation of the null geodesics.

\section{Asymptotic behavior of null geodesics}\label{B2}
In Ref.\cite{Amo:2021gcn}, the authors have studied the asymptotic behavior of the null geodesics in GR. They study the photons emitted outwards  at large $r$. Then they get the sufficient condition that the photons can be received by the observers at infinity. If the gravitational radiation is strong enough, this condition would be broken. Then the photons might not arrive at infinity. In this section, we will study the same thing in BD theory.
\subsection{The geodesic equations}\label{B3}
Here, we study the null geodesics in BD theory.
The geodesic equations can be transformed into the following two equations
\begin{equation}
r''=-\Gamma^r_{uu}u'u'-2\Gamma^r_{ur}u'r'-2\Gamma^r_{uA}u'(x^A)'-\Gamma^r_{rr}r'r'-2\Gamma^r_{rA}r'(x^A)'-\Gamma^r_{AB}(x^A)'(x^B)'\, ,
\end{equation}
and
\begin{equation}
u''=-\Gamma^u_{uu}u'u'-2\Gamma^u_{uA}u'(x^A)'-\Gamma^u_{AB}(x^A)'(x^B)'\, ,
\end{equation}
where $``\, '\, "$ denotes the derivative with respect to the affine parameter $\lambda$ of the null geodesic.
The related components of the Christoffel symbols are (ignore the lower order terms)~\cite{Barnich:2010eb}
\begin{eqnarray}
\Gamma^r_{uu}&=&\frac{\dot{\varphi}_1}{2\varphi_0r}-\frac{\dot{M}}{r}+\mathcal{O}\left(\frac{1}{r^{2}}\right)\, ,\nonumber\\
\Gamma^r_{ur}&=&\frac{M}{r^2}+\frac{\varphi_1}{2\varphi_0r^2}+\mathcal{O}\left(\frac{1}{r^{3}}\right)\, ,\nonumber\\
\Gamma^r_{uA}&=&-\frac{\partial_A M}{r}+\frac{1}{4r}\dot{c}_{AB}\eth_C\hat{c}^{BC}+\mathcal{O}\left(\frac{1}{r^{2}}\right)\, ,\nonumber\\
\Gamma^r_{rr}&=&\frac{\varphi_1}{\varphi_0r^2}+\mathcal{O}\left(\frac{1}{r^{3}}\right)\, ,\nonumber\\
\Gamma^r_{rA}&=&-\frac{\partial_A\varphi_1}{2\varphi_0r}+\frac{\eth_B\hat{c}^B{}_A}{2r}+\mathcal{O}\left(\frac{1}{r^{2}}\right)\, ,\nonumber\\
\Gamma^r_{AB}&=&\frac{1}{2}r\dot{c}_{AB}-rq_{AB}+\mathcal{O}\left(1\right)\,,\nonumber\\
\Gamma^u_{uu}&=&-\frac{1}{r}\frac{\dot{\varphi_1}}{\varphi_0}+\mathcal{O}\left(\frac{1}{r^{2}}\right)\, ,\nonumber\\
\Gamma^u_{uA}&=&-\frac{1}{r}\frac{\partial_A\varphi_1}{2\varphi_0}+\mathcal{O}\left(\frac{1}{r^{2}}\right)\, ,\nonumber\\
\Gamma^u_{AB}&=&rq_{AB}+\mathcal{O}\left(1\right)\, .
\end{eqnarray}
Since a future directed null geodesic is considered, we choose $u'>0$.  By these, the geodesic equations become
\begin{eqnarray}
\label{A5}
r''&=&\left(-\frac{\dot{\varphi}_1}{2\varphi_or}+\frac{\dot{M}}{r}\right)(u')^2
-2\left(\frac{M}{r^2}+\frac{\varphi_1}{2\varphi_0r^2}\right)u'r'
-2\left(-\frac{\partial_A M}{r}+\frac{1}{4r}\dot{c}_{AB}\eth_C\hat{c}^{BC}\right)u'(x^A)'\nonumber\\
&&-\frac{\varphi_1}{\varphi_0r^2}(r')^2-2\left(-\frac{\partial_A\varphi_1}{2\varphi_0r}+\frac{\eth_B\hat{c}^B{}_A}{2r}\right)r'(x^A)'
-\left(\frac{1}{2}r\dot{c}_{AB}-rq_{AB}\right)(x^A)'(x^B)'\, ,
\end{eqnarray}
and
\begin{equation}
\label{A6}
u''=\frac{1}{r}\frac{\dot{\varphi_1}}{\varphi_0}(u')^2+\frac{1}{r}\frac{\partial_A\varphi_1}{\varphi_0}u'(x^A)'
-rq_{AB}(x^A)'(x^B)'\, .
\end{equation}
Here and below, we ignore the lower order terms.
From the null condition of the tangent vector of null geodesics, i.e.,
\begin{equation}
\frac{\mathrm{d} s^{2}}{\mathrm{d}\lambda^2}=0=e^{2 \beta} \frac{V}{r} (u')^{2}-2 e^{2 \beta} u'r'+h_{A B}\left[ (x^{A})'-U^{A} u'\right]\left[ (x^{B})'-U^{B} u'\right]\, ,
\end{equation}
we get
\begin{equation}
\label{A7}
(u')^2=-2\left(1+\frac{2M}{r}\right)u'r'+(\eth_B\hat{c}^B{}_A) u'(x^A)'+[r^2q_{AB}+r(\hat{c}_{AB}+2Mq_{AB})](x^A)'(x^B)'\, ,
\end{equation}
or
\begin{equation}
\label{A8}
q_{AB}(x^A)'(x^B)'=\frac{1}{r^2}(u')^2+\frac{2}{r^2}u'r'-\frac{\eth_B\hat{c}^B{}_A}{r^2}  u'(x^A)'\,.
\end{equation}
So, from Eq.(\ref{A8}), one gets the relation between $u'$ and  $|(x^A)'|$ in leading order~\cite{Amo:2021gcn}:
\begin{equation}
\label{A9}
u'=\big[r+\mathcal{O}\left(1\right)\big]|(x^A)'|\, ,
\end{equation}
where $|(x^A)'|$ is defined as
$$|(x^A)'|=\sqrt{q_{AB}(x^A)'(x^B)'}\, .$$
With the initial condition $r'=0$, the second derivative of $r$ is
\begin{equation}
r''=r\Big(q_{AB}-\frac{1}{2}\dot{\hat{c}}_{AB}+\dot{M}q_{AB}\Big)(x^A)'(x^B)'\equiv r\Omega_{AB}(x^A)'(x^B)'\, .
\end{equation}
Therefore, the tensor components
\begin{equation}
\label{A10}
\Omega_{AB}=q_{AB}-\frac{1}{2}\dot{\hat{c}}_{AB}+\dot{M}q_{AB}=q_{AB}(1+\dot{M})+\frac{1}{2}N_{AB}
\end{equation}
will determine the behavior of photons at infinity. Obviously, if
\begin{equation}
\label{A20}
\Omega_{AB}(x^A)'(x^B)'\geqslant0,
\end{equation}
then $r''\geqslant0$ at $\lambda=0$, and then $r'\geqslant0$ for all $\lambda>0$. The reason is given as follows: If $r'(\lambda_c)=0$ for $\lambda_c>0$, and $r'(\lambda)<0$ for $\lambda>\lambda_c$, then $r''(\lambda_c)<0$. This is contradict with $r''(\lambda_c)=r\Omega_{AB}(x^A)'(x^B)'\geqslant0$.
In fact, this conclusion holds for any null geodesics with $r'(0) \geqslant 0$.

Therefore, $r$ is increasing along the geodesic.
It is not hard to find that the condition (\ref{A20}) is equivalent to that the  two eigenvalues of $\Omega_{AB}$, $k_1$ and $k_2$, are non-negative.

\subsection{Behavior of $r(\lambda)$ and $u(\lambda)$}\label{B4}
In this subsection, we will study the asymptotic behavior of $r(\lambda)$ and $u(\lambda)$. Substituting Eq.(\ref{A7}) into Eq.(\ref{A5}),  and considering Eq.(\ref{A9}), we obtain
\begin{eqnarray}
\label{A17}
r''&=&\frac{1}{r}\left(-2\dot{M}+\frac{\dot{\varphi_1}}{\varphi_0}\right)u'r'+r\Omega_{AB}(x^A)'(x^B)'
+\frac{1}{r}\left(\frac{\partial_A\varphi_1}{\varphi_0}-\eth_B\hat{c}^B{}_A\right)r'(x^A)'-\frac{1}{r^2}\frac{\varphi_1}{\varphi_0}(r')^2\nonumber\\
&&+\frac{1}{r}\left[2\partial_AM-\frac{1}{2}\dot{c}_{AB}\eth_C\hat{c}^{BC}+\left(\dot{M}-\frac{\dot{\varphi}_1}{2\varphi_0}\right)\eth_B\hat{c}^B{}_A\right]u'(x^A)'\nonumber\\
&=&\frac{1}{r}\left(-2\dot{M}+\frac{\dot{\varphi_1}}{\varphi_0}\right)u'r'+r\Omega_{AB}(x^A)'(x^B)'-\tilde{C_1}\frac{1}{r^2}(r')^2\, ,
\end{eqnarray}
where $\tilde{C_1}$ is a constant,  and lower order terms  have been omitted. So if
\begin{eqnarray}
\label{A11}
-2\dot{M}+\frac{\dot{\varphi_1}}{\varphi_0}&\geqslant & 0\, ,\nonumber\\
\Omega_{AB}(x^A)'(x^B)' &\geqslant &0\, ,
\end{eqnarray}
then $r''\geqslant -\tilde{C_1}(r')^2/r^2$. Solving this inequality, we get
\begin{equation}
r\geqslant \tilde{C_2}\lambda+\tilde{C_3}\, ,
\end{equation}
where $\tilde{C_2}$, $\tilde{C_3}$ are constants,  and $\tilde{C_2}$ is positive. This means that $r$ approaches to infinity as $\lambda$ becomes infinity. So the photons can escape to the infinity.

In order to know whether the photons can escape to the infinity in a finite time, we have to calculate $u''$. Substituting Eq.(\ref{A8}) into Eq.(\ref{A6}),  and  using Eq.(\ref{A9}), we get
\begin{equation}
\label{A18}
u''=\Big(\frac{\dot{\varphi}_1}{\varphi_0}-1\Big)\frac{1}{r}(u')^2-\frac{2}{r}u'r'\, .
\end{equation}
If $\dot{\varphi}_1/\varphi_0-1\leqslant 0$, then $u''\leqslant -2u'r'/r$. This implies $0\leqslant u'\leqslant\tilde{C}_4 r^{-2}$, where $\tilde{C_4}$ is a positive constant.
Because $r$ has the order $\mathcal{O}\left(\lambda \right)$, $u$ is finite as $\lambda$ approaches to infinity~\cite{Amo:2021gcn}. So the photons can be received by the observers at infinity in a finite time.

In conclusion, the photons could arrive at the future null infinity if the following conditions are satisfied, i.e.,
\begin{eqnarray}
\label{A12}
\Omega_{AB}(x^A)'(x^B)'\geqslant0\, ,\nonumber\\
2\dot{M}\leqslant\frac{\dot{\varphi}_1}{\varphi_0}\leqslant 1\, .
\end{eqnarray}
The second condition above is obviously absent in  GR. Of course, the details of the tensor $\Omega_{AB}$ are also different from the one in GR.

\subsection{The sufficient condition}\label{B5}
We now turn to look for the meaning of Eq.(\ref{A20}). This condition is equivalent to that the two eigenvalues of $\Omega_{AB}$ are non-negative.
This suggests
$$\dot{M}+1\pm \sqrt{\frac{1}{8}N_{AB}N^{AB}}\geqslant0\, ,$$
or
\begin{eqnarray}
\label{A13}
\dot{M}+1&\geqslant & 0 \,,\nonumber\\
(\dot{M}+1)^2&\geqslant & \frac{1}{8}N_{AB}N^{AB}\, .
\end{eqnarray}
From Eq.(\ref{A4}), we know
\begin{equation}
\frac{1}{8}N_{AB}N^{AB}=-\dot{M}-\frac{1}{4} \eth_A \eth_B N^{A B}-\frac{2 \omega+3}{4}\left(\frac{\dot{\varphi}_1}{\varphi_{0}}\right)^{2}\, .
\end{equation}
So inequalities (\ref{A13}) becomes
\begin{equation}
\dot{M}\geqslant\max \Bigg{\{}-1\, ,~~ -\frac{3}{2}+\frac{1}{2}\sqrt{5-\eth_A \eth_B N^{A B}-(2 \omega+3)\left(\frac{\dot{\varphi}_1}{\varphi_{0}}\right)^{2}} \Bigg{\}}\, .
\end{equation}
Combining inequalities (\ref{A12}), the sufficient condition that the photons emitted at large $r$ with $r'\geqslant0$ can arrive at the future null infinity is
\begin{equation}
\label{A14}
\max \Bigg{\{}-1\, ,~~ -\frac{3}{2}+\frac{1}{2}\sqrt{5-\eth_A \eth_B N^{A B}-(2 \omega+3)\left(\frac{\dot{\varphi}_1}{\varphi_{0}}\right)^{2}} \Bigg{\}}\leqslant\dot{M}\leqslant\frac{\dot{\varphi}_1}{2\varphi_0}\leqslant\frac{1}{2}\, .
\end{equation}
The term $\eth_A \eth_B N^{A B}$ does not vanish in general. Actually, it will be vanished if and only if the news tensor is vanished (see the detailed proof by Ashtekar {\it et. al} in~\cite{Ashtekar:1981hw,Ashtekar:2014zsa}).

The geometric meaning of the term $\eth_A \eth_B N^{A B}$ can be understood as follows: Assuming there exists a manifold $\tilde{M}$ with a boundary $\mathcal{I}$ equipped with a metric $\tilde{g}_{ab}=\Omega^2 g_{ab}$ and a conformal transformation from $M$ onto $\tilde{M} \backslash \mathcal{I}$, where $(M, g_{ab})$ is the physical spacetime, and considering the null normal vector
\begin{equation}
n_a\equiv \nabla_a \Omega=\nabla_a \Big(\frac{1}{r}\Big)=-\frac{1}{r^2}(dr)_a\, ,
\end{equation}
the leading order of Weyl tensor $\tilde{C}_{abc}{}^{d}$ on $\tilde{M}$ is given by
\begin{equation}
K_{abc}{}^{d}=\Omega^{-1} \tilde{C}_{abc}{}^d\,,
\end{equation}
and the magnetic part of $K_{abc}{}^d$ has a form
\begin{equation}
{}^*K^{ac}={}^*K^{abcd}n_b n_d\,,
\end{equation}
where ${}^*K_{abcd}$ is the dual of $K_{abcd}$.
It is not hard to find the divergence of news tensor is proportional to the leading order magnetic part of the Weyl tensor of the spacetime.
By defining the current $j_A= \,^*K^0{}_A$, the relation of $j_A$ and $\eth_A \eth_B N^{A B}$ is given by
\begin{equation}
\eth_A \eth_B N^{A B}
=-2\epsilon^{AB}\eth_A j_B \, ,
\end{equation}
where $\epsilon_{AB}$ is the component of the Levi-Civita tensor on the standard two dimensional sphere. So the term $\eth_A \eth_B N^{A B}$ has a clear geometric meaning and
does not vanish in general.

To get detailed information from the inequalities in the above, for example, inequalities (\ref{A14}), we have to know the details of $\eth_A \eth_B N^{A B}$.
This can not be achieved without the information on the magnetic part the leading order of the Weyl tensor.  However,  by considering an inequality
\begin{equation}
\label{Inequality1}
\sqrt{5-x}\leqslant -\frac{\sqrt{5}}{10}x+\sqrt{5}\, ,
\end{equation}
the contribution of $\eth_A \eth_B N^{A B}$ does not appear in the inequality after integrating on the sphere. This point can be found as follows.
Inequality (\ref{Inequality1}) implies  that inequalities (\ref{A14}) can be replaced by
\begin{equation}
\max \Bigg{\{} -1\, ,~~\frac{\sqrt{5}-3}{2}-\frac{\sqrt{5}}{20}\Bigg[\eth_A \eth_B N^{A B}+(2 \omega+3)\left(\frac{\dot{\varphi}_1}{\varphi_{0}}\right)^{2}\Bigg] \Bigg{\}}
\leqslant\dot{M}\leqslant\frac{\dot{\varphi}_1}{2\varphi_0}\leqslant\frac{1}{2}\, .
\end{equation}
Integrating the above inequality on the sphere, and  multiplying the result with $\varphi_0/4\pi$, we get
\begin{equation}
\label{A15}
\max\Bigg{\{}-\varphi_0,\quad  \frac{\sqrt{5}-3}{2}\varphi_0-\frac{2\omega+3}{16\sqrt{5}\pi}\varphi_0\int\sqrt{q}\left(\frac{\dot{\varphi}_1}{\varphi_0}\right)^2d\Omega\Bigg{\}}
\leqslant\dot{m}\leqslant\frac{1}{8\pi}\int\sqrt{q}\dot{\varphi_1}d\Omega\leqslant\frac{1}{2}\varphi_0\, ,
\end{equation}
where
\begin{equation}
m=\frac{\varphi_0}{4\pi}\int M\sqrt{q}d\Omega
\end{equation}
is the Bondi mass of the asymptotically flat system~\cite{Hou:2020tnd}. From Eq. (\ref{A4}), we know $\dot{m}$ can not be positive.   Remarkably, the term in the inequalities (\ref{A15}), i.e.,
\begin{equation}
J_\varphi\equiv\frac{2\omega+3}{16\pi}\varphi_0\int\sqrt{q}\left(\frac{\dot{\varphi}_1}{\varphi_0}\right)^2d\Omega
\end{equation}
represents the energy flux of the scalar gravitational waves~\cite{Hou:2020tnd}. From the inequalities (\ref{A14}), we know $-2\leqslant\dot{\varphi}_1/\varphi_0\leqslant1$. This gives
\begin{equation}
J_\varphi\leqslant(2\omega+3)\varphi_0\, .
\end{equation}
Here, $\omega$ is assumed to be positive. Therefore, when  $$J_\varphi<(5-\sqrt{5})\varphi_0/2\, ,$$ i.e.,  the scalar gravitational radiation is weak, we have
\begin{equation}
\frac{\sqrt{5}-3}{2}\varphi_0-\frac{1}{\sqrt{5}}J_\varphi
\leqslant\dot{m}\leqslant \frac{1}{2}\dot{Q}\leqslant\frac{1}{2} \varphi_0\, ,
\end{equation}
where
\begin{equation}
Q=\frac{1}{4\pi}\int\sqrt{q}\varphi_1d\Omega\,.
\end{equation}
However, when $$(5-\sqrt{5})\varphi_0/2\leqslant J_\varphi\leqslant (2\omega+3)\varphi_0\, ,$$ i.e., the scalar gravitational radiation is strong enough, we obtain
\begin{equation}
-\varphi_0\leqslant\dot{m}\leqslant\frac{1}{2}\dot{Q}\leqslant\frac{1}{2}\varphi_0\, .
\end{equation}
Therefore, the range of $\dot{m}$ has a closed relation to the value of the scalar at infinity and the scalar gravitational radiation.

The upper bound is decided by $\dot{Q}$. When $\dot{Q}<0$, the upper bound is proportional to $\dot{Q}$.
However, the lower bound is decided by $J_\varphi$. When $J_\varphi$ is small, the lower bound  decreases as $J_\varphi$ increases. But when $J_\varphi$ is large, the lower bound is a constant.
The relation between the lower bound and $J_\varphi$ can be found in Fig.1, while the relation between upper bound and $\dot{Q}$ is depicted in Fig.2.
Besides, to ensure the photons to arrive at infinity, the scalar gravitational radiation can not be too large.

\begin{figure}[htb]
\label{lower1}
  \centering
  \includegraphics[width=10cm]{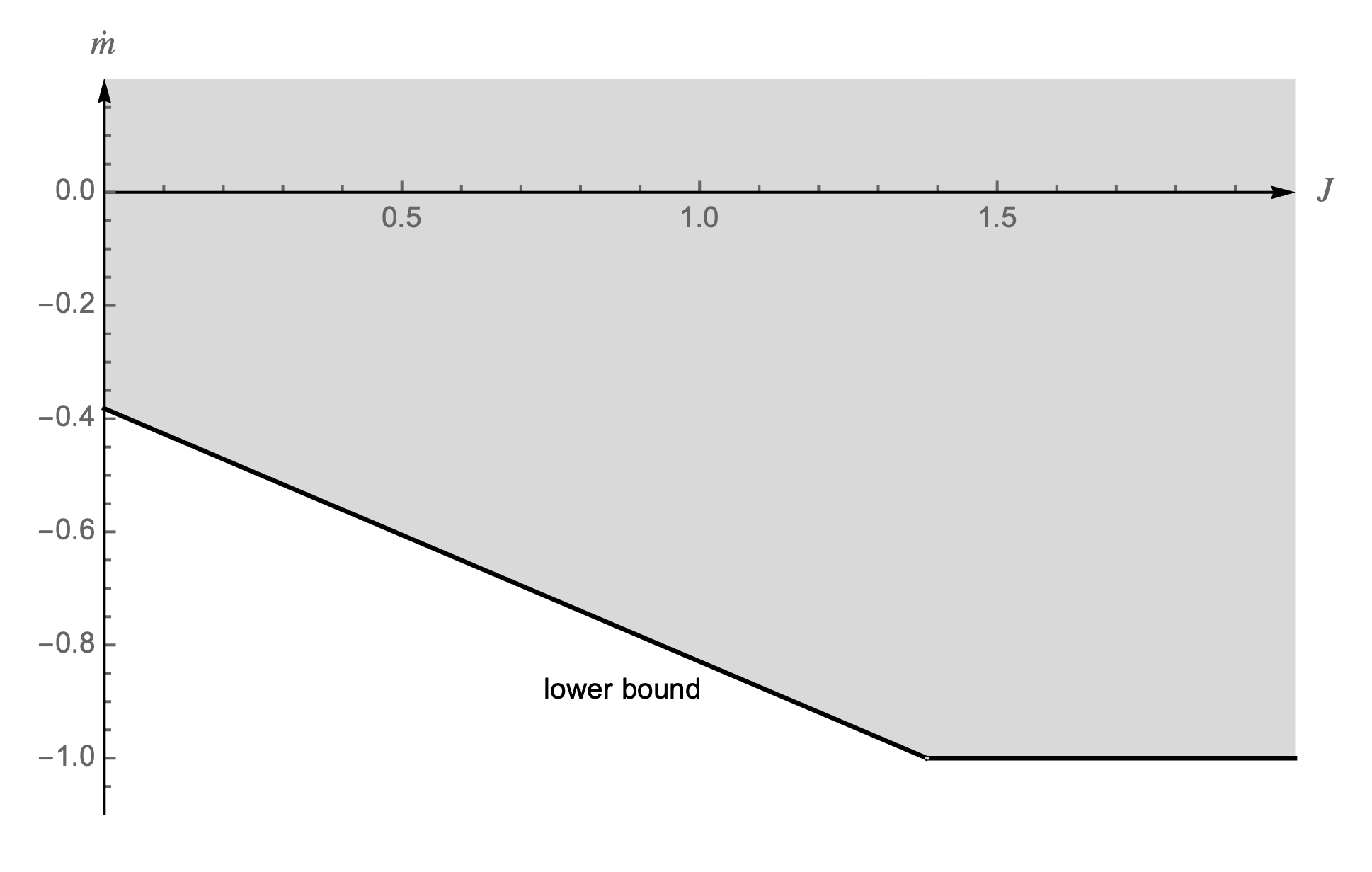}
  \caption{The lower bound with respect to $J=J_\varphi/\varphi_0$.}
  \end{figure}

\subsection{The case in GR}\label{B6}
When $\varphi_1=0,\varphi_0=1$, the results in subsection \ref{B5} reduce to the case in GR. The sufficient condition (\ref{A14}) becomes
\begin{equation}
\max \Bigg{\{}-1\, ,~~ -\frac{3}{2}+\frac{1}{2}\sqrt{5-\eth_A \eth_B N^{A B}} \Bigg{\}} \leqslant \dot{M}\leqslant0\, ,
\end{equation}
and the condition  (\ref{A15}) reduces to
\begin{equation}
\frac{\sqrt{5}-3}{2} \leqslant \dot{m}\leqslant0\, .
\end{equation}
In GR, $\dot{m}\leqslant0$ is satisfied automatically. So this means the mass loss of the system can not be too large.
If the gravitational radiation is strong enough, one may not receive the photons with $r'\geqslant0$ at infinity. The lower bound of $\dot{m}$ is given by
\begin{equation}
\label{bound}
b=\frac{\sqrt{5}-3}{2}\frac{c^3}{G}= -0.3820~\frac{c^3}{G}=-0.3820~\frac{m_p}{t_p}\, ,
\end{equation}
where the Newtonian gravitational constant $G$ and   the speed of light $c$ in vacuum have been restored, and
$$m_p=\sqrt{\frac{\hbar c}{G}}\, ,\qquad t_p = \sqrt{\frac{\hbar G}{c^5}} $$
are Planck mass and Planck time respectively. The value of $b$ in Eq.(\ref{bound}) is $-1.546\times10^{35}\mathrm{kg}/\mathrm{s}$.
This is a very loose bound. Intuitively, if the bound is saturated, the system will lose a solar mass $M_{\odot}$ in $1.28\times 10^{-5}\mathrm{s}$. With this rate of mass loss, a super massive black hole
with mass $10^8 M_{\odot}$ will lose all of its mass within  half an hour.

Comparing with GR, we can find the scalar field $\varphi$ decreases the lower bound in BD when  $\varphi_0$ is on the order of one (for example,  $\varphi_0=1$).
When $J_\varphi=0$, i.e.,  the scalar gravitational radiation is absent, the range of $\dot{m}$ in BD reduces to that in GR. But as $J_\varphi$ increases, the lower bound in BD decreases. When $J_\varphi\geqslant(5-\sqrt{5})\varphi_0/2$, the lower bound is $2.62b$.
The upper bound in BD is different from the one in GR. When $\dot{Q}>0$, the upper bound in BD is zero. This is the same as the upper bound in GR. However, when $\dot{Q}<0$, the upper bound in BD is less than zero, and it is proportional to $\dot{Q}$. This tells us the mass loss can not be too slow and too fast if $\dot{Q}<0$.

If $\varphi_0\neq1$, the behavior of the upper bound is the same as  the case with $\varphi_0=1$. But the lower bound is different, even when $J_\varphi$ is vanished. Besides $J_\varphi$, the lower bound is also affected by the effective gravitational constant in the infinity, i.e.,
\begin{equation}
G_0=\frac{G}{\varphi_0}\, .
\end{equation}
So the role of $\varphi_0$ is nothing but the effective Newtonian constant.

\begin{figure}[htb]
\label{upper1}
  \centering
  \includegraphics[width=10cm]{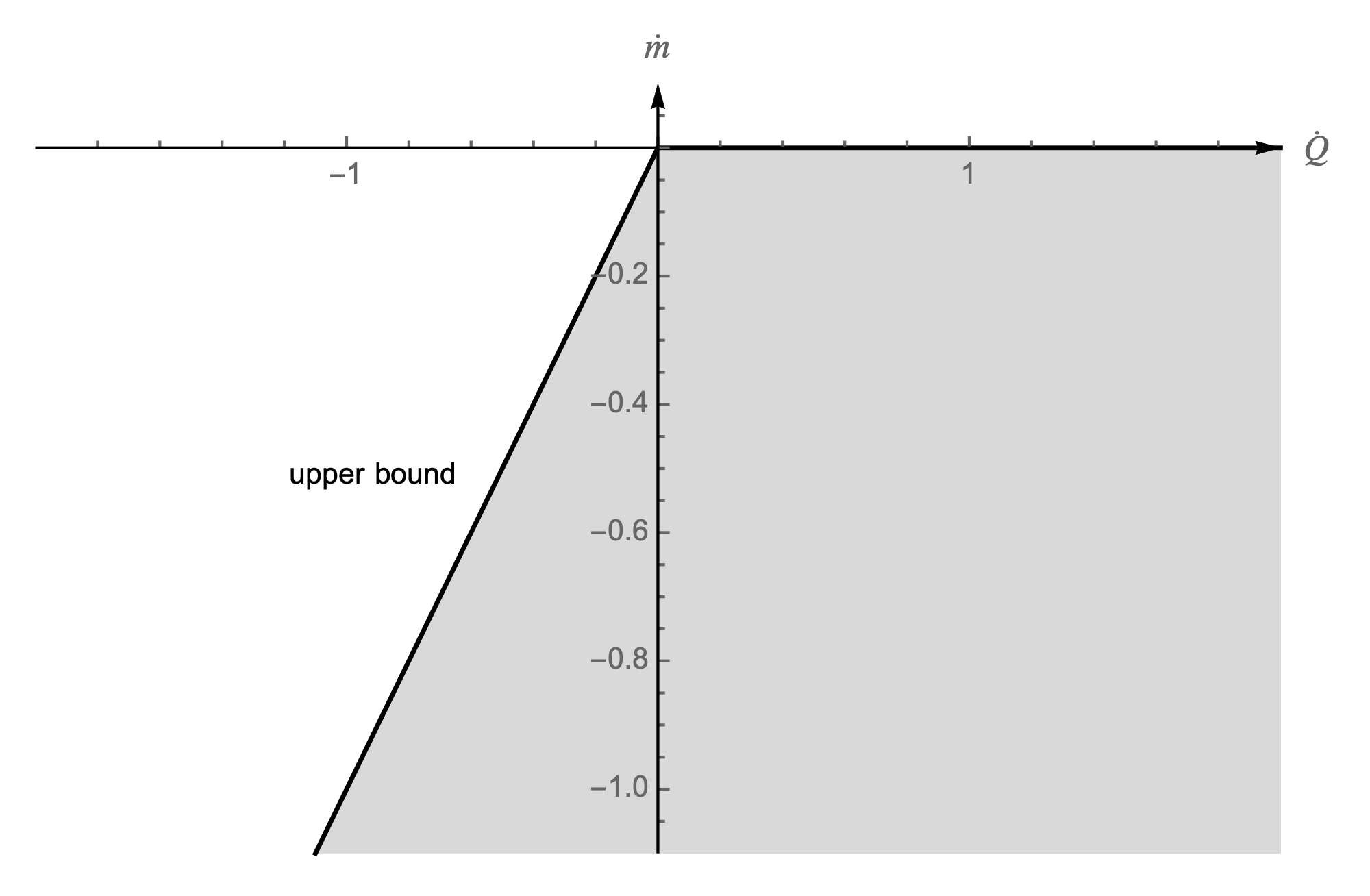}
  \caption{The upper bound with respect to $\dot{Q}$.}
\end{figure}

\subsection{The case in spherically symmetric spacetime}\label{B7}
In GR, the metric of a spherically symmetric spacetime in the Bondi-Sachs coordinates has a form
\begin{equation}
\mathrm{d} s^{2}=e^{2 \beta} \frac{V}{r} \mathrm{d} u^{2}-2 e^{2 \beta} \mathrm{d} u \mathrm{d} r+q_{A B}\mathrm{d} x^{A}\mathrm{d} x^{B}.
\end{equation}
Solving the equations of motion (\ref{A1}), (\ref{A2}), and (\ref{A16}), we get the solution in vacuum:
\begin{eqnarray}
\beta&=&0\, ,\nonumber\\
V&=&2M-r\, ,\nonumber\\
\dot{M}&=&0\, .
\end{eqnarray}
This is nothing but the Schwarzchild spacetime. The result is consistent with the well-known Birkhoff theorem.
Now, Eqs. (\ref{A17}) and (\ref{A18}) become
\begin{eqnarray}
r''&=&r q_{AB}(x^A)'(x^B)'-\tilde{C_5}\frac{1}{r^2}(r')^2\, ,\\
u''&=&-\frac{1}{r}(u')^2-\frac{2}{r}u'r'\, ,
\end{eqnarray}
where $\tilde{C_5}$ is a constant. After repeating the calculation in subsection \ref{B3} and \ref{B4}, we know the photons with $r'\geqslant0$ can escape to the infinity without any  constraints.

However, in BD, things are different. By considering the determinant condition (\ref{A3}), the metric can be written as
 \begin{equation}
\mathrm{d} s^{2}=e^{2 \beta} \frac{V}{r} \mathrm{d} u^{2}-2 e^{2 \beta} \mathrm{d} u \mathrm{d} r+r^2 \frac{\varphi_0}{\varphi} q_{A B}\mathrm{d} x^{A}\mathrm{d} x^{B}\, .
\end{equation}
So the solution of the equations of motion are
\begin{eqnarray}
\beta &=&-\frac{\varphi_1}{2\varphi_0}\frac{1}{r}+\mathcal{O}\left(\frac{1}{r^{2}}\right)\, ,\nonumber\\
V&=&-r+2M+\mathcal{O}\left(\frac{1}{r}\right)\, ,\nonumber\\
\dot{M}&=&-\frac{2 \omega+3}{4}\left(\frac{\dot{\varphi}_1}{\varphi_{0}}\right)^{2}+\mathcal{O}\left(\frac{1}{r}\right)\, .
\end{eqnarray}
Repeating the calculation in subsections \ref{B3}, \ref{B4},  and \ref{B5}, we get the sufficient condition
\begin{equation}
-1\leqslant-\frac{2\omega+3}{4}\left(\frac{\dot{\varphi}_1}{\varphi_{0}}\right)^{2}\leqslant\frac{\dot{\varphi_1}}{2\varphi_0}\leqslant\frac{1}{2}\, .
\end{equation}
This result is consistent with inequality (\ref{A14}), and can be transformed into a form
\begin{equation}
\label{IEQ1}
0\leqslant\frac{\dot{\varphi}_1}{\varphi_{0}}\leqslant\min\Bigg{\{}1,~\frac{2}{\sqrt{2\omega+3}}\Bigg{\}}\, ,
\end{equation}
or
\begin{equation}
\label{IEQ2}
-\frac{2}{\sqrt{2\omega+3}}\leqslant\frac{\dot{\varphi}_1}{\varphi_{0}}\leqslant-\frac{2}{2\omega+3}\, .
\end{equation}
In spherically symmetric spacetime of BD theory, $J_\varphi$ is equal to the rate of the mass loss of the system~\cite{Hou:2020tnd}, i.e.,
\begin{equation}
\dot{m}=-J_\varphi\, .
\end{equation}
Integrating the inequalities (\ref{IEQ1}) and (\ref{IEQ2})  over the two dimensional sphere,  then we find the sufficient condition
\begin{equation}
\dot{\varphi}_1\geqslant0,\qquad \max\Bigg{\{}-\varphi_0\, ,~~-\frac{2\omega+3}{4}\varphi_0\Bigg{\}}\leqslant\dot{m}\leqslant0\, ,
\end{equation}
or
\begin{equation}
\dot{\varphi}_1<0,\qquad -\varphi_0\leqslant \dot{m}\leqslant-\frac{\varphi_0}{2\omega+3}\, .
\end{equation}
Therefore, in the spherically symmetric spacetimes, due to the scalar field, the situation in BD is very different from the one in GR. In GR, the photons can escape to the infinity without any conditions.
However, there is a sufficient condition of $\dot{m}$ to ensure the photons to arrive at infinity in spherically symmetric spacetime in BD. This is because the scalar gravitational radiation affects the behavior of the photons in some sense, whereas there is no gravitational radiation in any spherically symmetric spacetime in GR.

\section{conclusions and discussion}
\label{ConDis}
In this paper, by using the  Bondi-Sachs formalism, we have studied the asymptotic behavior of the future null geodesics in BD. We get the sufficient condition that the photons emitted at large $r$ with $r'\geqslant0$ can arrive at infinity.
In GR,  the arrival of the photons to the infinity  suggests that the  Bondi mass loss can not be too fast. The upper bound of $|\dot{m}|$ is $0.3820c^3/G$. In BD, due to the existence of the scalar field $\varphi$, the lower bound is decreased with respect to the flux of the scalar gravitational radiation.
In addition, $\dot{m}$ can not be arbitrary and has an upper bound  when $\dot{\varphi}_1$ is negative.
So there is also a possibility that the photons can not arrive at infinity. The situation is similar to the case in GR.

As we know, the photons can escape to the infinity in the flat spacetime.
Therefore,  the photons emitted outwards at large $r$ are expected to have the same behavior because the metric there is nearly flat.
However, the work in \cite{Amo:2021gcn} indicates the asymptotically flat spacetime is not as simple as we thought.
When the gravitational radiation is  intense enough, the photons may not reach the infinity.
This suggests there may be a maximum luminosity in the spacetime to make sure the photons to arrive at infinity.
It should be pointed out here that the statement  in this paper does not provide  a proof of the Dyson's maximum luminosity. We
just provide a clue for the existence of  Dyson's maximum luminosity.

In GR, from the bound of the rate of the mass loss, we can find that the luminosity, $P$, of the asymptotically flat spacetime has a maximum value $P_{\mathrm{m}}$, i.e.,
\begin{equation}
P=|\dot{m}|c^2\leqslant P_{\mathrm{m}} = 0.3820 P_*\, ,
\end{equation}
where $P_*=c^5/G$ is the so-called ``one Dyson unit". This maximum luminosity is first proposed by Dyson  long time ago~\cite{Dayson1963, Barrow:2017atq}. By considering the radiation of a  binary star system, he got a maximum luminosity $P_{\mathrm{m}} =(125/8) P_*$.
In fact,  a lot of works have suggested  that there is really a maximum luminosity for any kinds of radiation.
For example, numerical relativity simulations of critical collapse yield a  tighter bound, i.e.,  $P_{\mathrm{m}} \approx 0.2 P_*$.  However, to get this value, a spherical symmetry
has been assumed in the  simulations~\cite{Cardoso:2018nkg}.
Recently, Jowsey and Visser have studied the bound in Vaidya spacetime and an evaporating version of Schwarzschild's  constant density star~\cite{Jowsey:2021gny}.
Then they found  some additional conditions are necessary to get a bounded luminosity. Otherwise, the luminosity can be arbitrarily large.  Some related discussion on this topic
can be found in~\cite{Faraoni:2021wre,Schiller:2021ois,Faraoni:2021sep}.
All of these analyses are based on some specific physical processes, and most of them are in spherically symmetric spacetimes.
Obviously, the bound in present paper does not contradict  to all of bounds founded in these literatures.
However, the logic of this paper is different from theirs.  The starting point of our discussion is the light  influenced under the gravitational radiation or gravitational wave.
To ensure the photons, emitted outwards in the region where the metric is nearly flat, can arrive at infinity, the luminosity can not be too large.
The analysis is performed near the infinity of the spacetime, so the details of the  physical processes deep inside the spacetime are not necessary.
For this reason, our result is model-independent in framework of GR.  Finally, in BD, we have to consider the contribution from the scalar radiation, and the maximal luminosity has to be enlarged to one Dyson unit $P_*$.

The lower bound of $\dot{m}$ is roughly equal to lose a Planck mass $m_p$ in a Planck time $t_p$.
It is so large that such violent astronomical phenomena can not happen.
In the case of weak field and low velocity, the energy loss of the binary star systems is given by
\begin{equation}
\label{dEdt}
\frac{\mathrm{d}E}{\mathrm{d}t}\sim - \left(\mathcal{M}\omega\right)^\frac{10}{3} \cdot \frac{G^{\frac{7}{3}}}{c^5}\,,
\end{equation}
where $\mathcal{M}$ is the chirp mass of the binary system and $\omega$ is the frequency of the gravitational waves.
So, for a binary star system, assuming the formula (\ref{dEdt}) can be naively extrapolated to very high frequency (it is not suitable because  the low velocity condition is broken, and post Newtonian approximation has to be considered),
when the photons emitted at large $r$ with $r'\geqslant0$ can not escape to the infinity, the orbit frequency and the frequency of the gravitational waves are very high.
Actually, when the bound $b$ is saturated, i.e., $\mathrm{d}E/\mathrm{d}t\sim b c^2$, we have
$$\omega \sim \frac{1}{t_p}\cdot \Big(\frac{m_p}{\mathcal{M}}\Big) \sim 10^{5}\Big( \frac{M_{\odot}}{\mathcal{M}} \Big)\mathrm{Hz}\, .$$
This simple investigation implies  that  the geometrical optics approximation is still valid for the binary gravitational system with an astronomical chirp mass.  So the  model of null geodesics can be used despite of the huge gravitational radiation of the system.
Certainly, in the case $\mathcal{M}$ is less than $ 10^{-9}M_{\odot}$, the frequency of the gravitational waves is larger than $\omega\sim 10^{14} \mathrm{Hz}$, i.e., the frequency of  visible light.
In this case, we have to consider the full Bondi-Sachs formalism  in GR or BD with the electromagnetic field, and find the
condition that electromagnetic waves can arrive at the null infinity.   This kind of study may reveal the influence of the gravitational waves to the electromagnetic waves in a global way.
These need further study.

Although the phenomenon of this kind of fast mass loss may not be observed astronomically, it might occur when a tiny black hole was created in a very high energy experiment in laboratory.
If the system can be treated in a classical way, our results suggest  that we may not receive some  photons emitted far away from this tiny black hole if its mass loss is too fast.
Needless to say,  the quantum effect will be dominant in this case. So this classical model will be failed  in this extreme gravitational system, especially, in the cases where the effect of quantum gravity  is involved.

It has been shown that the range of $\dot{m}$ in BD is very different from the one in GR, especially in the case with a spherical symmetry.
Probably, the lower bound of the mass loss could impose some restrictions on a given gravity theory.
So, in other gravity theories, it is interesting to study whether there is a realistic lower bound of $\dot{m}$ which could has some  astronomically observable effect.

\section*{Acknowledgement}

 We would like to thank Gary Gibbons for his valuable communication on the maximum luminosity.
This work was supported in part by the National Natural Science Foundation of China with grants No.11622543,
No.12075232, No.11947301, and No.12047502. This work is also supported by the Fundamental Research Funds for the
Central Universities under Grant No: WK2030000036, and the Key Research Program of the Chinese Academy of Sciences, Grant NO. XDPB15.

\end{document}